# Near real-time channel selection for Distributed Acoustic Sensing technology

Emanuele Bozzi,[1] Giulio Pascucci,[1] Giacomo Rapagnani,[1] Gian Maria Bocchini,[2] Rebecca Harrington,[2] Arantza Ugalde,[3] Gilberto Saccorotti,[4] Francesco Grigoli[1]

[1] Department of Earth Sciences, University of Pisa, Pisa, Italy
[2] Institute of Geosciences, Ruhr University Bochum, Bochum, Germany
[3] Department of Marine Geosciences, Institut de Ciències del Mar, CSIC, 08003 Barcelona, Spain
[4] Istituto Nazionale di Geofisica e Vulcanologia, Pisa, Italy

*Corresponding author: emanuele.bozzi@dst.unipi.it, emanuele.bozzi96@gmail.com

## ABSTRACT

*Distributed Acoustic Sensing (DAS) technology is advancing seismic monitoring by providing dense observations near earthquake sources. However, the resulting data volumes often limit real-time processing capability, with most seismological applications focusing on retrospective analysis of seismic sequences. To address this challenge, we introduce ORION, a fast and versatile selector of high-quality DAS channels that efficiently reduces the amount of data to analyze. The method first adopts spatial clustering to identify cable segments with similar geometrical attributes (e.g, azimuth), and then performs channel selection within each section using waveform attributes (e.g., signal-to-noise ratio); this approach enables spatial sub-sampling while preserving azimuthal coverage. We demonstrate the flexibility of the selector across several cable geometries. Finally, we analyze a seismic sequence using ORION-selected channels and compare the source locations with those from a more conventional uniform distribution of channels along the cable, showing improvements in hypocenter accuracy.*

## INTRODUCTION

Distributed Acoustic Sensing (DAS) is an emerging data acquisition technology that transforms conventional telecommunication, or built-in fiber-optic cables, into ultra-dense arrays of seismic sensors measuring strain/strain-rate [1–4]. In seismological applications, DAS systems can fill monitoring gaps in poorly instrumented regions [5–9] and complement existing seismic networks [10, 11], producing massive and heterogeneous datasets [12]. The full exploitation of DAS data density allows the adoption of innovative approaches that improve monitoring procedures [13–22]. However, such new techniques remain largely limited to offline analyses due to the lack of efficient processing strategies capable of handling high data complexity both in terms of size and signal quality [23]. Although data compression strategies help reduce storage demands [24], thereby mitigating the long-term impact of DAS experiments, rapid processing workflows still rely on spatial sub-sampling and stacking of nearby channels [25, 26]. Furthermore, joint analyses of seismic events using conventional networks and fiber-optic sensors are challenged by the highly unbalanced number of observation points [8, 23], making spatial sub-sampling of DAS data/channels essential.

A common practice to manage massive DAS datasets in near real-time involves the tailored selection of a subset of channels, which are then integrated into conventional seismic data processing algorithms in a second step. One possible approach is to focus on waveform attributes, identifying good-quality channels that can help manage the complex spatio-temporal variations of signal properties (e.g., caused by coupling or the uniaxial sensitivity of the sensor) [27, 28]. For example, Rodriguez et al. (2025) introduced an efficient automatic selection method based on a machine-learned channel quality index that integrates multiple waveform attributes (e.g., signal mean, median, variance, kurtosis). On the other hand, to perform seismic event location with DAS sensors, channel selection must also account for network configuration [29], as the final set of channels should ideally provide sufficient azimuthal coverage [30, 31]. Moreover, the number of sensors within azimuthal wedges of a given width should remain roughly constant to avoid over- or under-sampling along specific directions and thus bias source location. A common strategy to deal with the above criteria is to select traces at fixed intervals along the cable, though this does not guarantee the exclusion of poor-quality channels. A possible alternative is to relax purely waveform attribute-based selection and allow for the inclusion of suboptimal channels (relatively good quality but not the overall best quality) to accommodate the azimuthal requirements.

Here, we introduce ORION, an automatic DAS channel selector that combines waveform attributes and geometry information by jointly balancing data quality and spatial coverage in a single workflow (Fig. 1). ORION is designed to meet the following requirements:

- Being robust to varying types of cable geometry,

- Ensuring sufficient spatial coverage along the cable, including a maximum distance between selected channels,

- Allowing users (e.g., seismologists) to adjust the relative weighting of waveform attributes and spatial coverage, enabling them to focus on either high-quality channels or azimuthal coverage,

- Supporting the retrieval of a defined number of channels, e.g., for a successive integration of DAS data with existing conventional networks,

- Operating in near real-time.

We evaluate the performance of ORION in two steps: (i) an application to several DAS cable geometries to test its versatility, and (ii) an analysis of its impact on earthquake location procedures. In the latter step, we compare ORION-selected traces with a spatially uniform sub-sampling strategy. The results show that integrating waveform-driven and geometry-based criteria offers support for both near-real-time signal evaluation and more effective seismic monitoring.

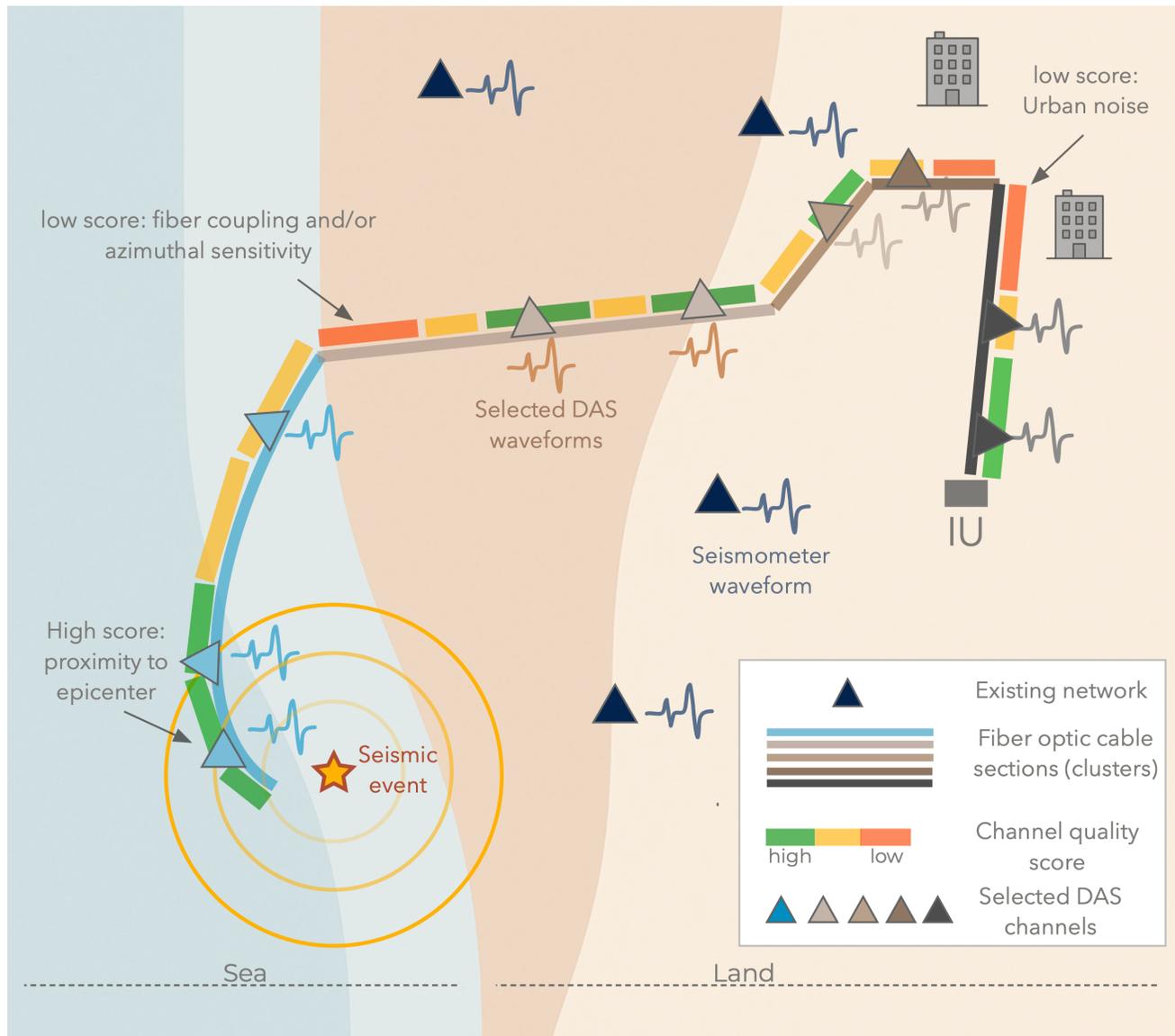

**Figure 1:** Sketch of the proposed automatic DAS channel selector for a hybrid land–marine cable deployment (colored line) with offshore seismicity and an onshore seismic network (dark blue triangles). The selection procedure is schematized as follows: (1) Spatial clustering, illustrated by different colors of the cable; (2) Channel quality score, simplified with three colors, red, yellow, and green; (3) Final channel selection, identified by triangles with subsections enforcing a minimum interstation distance in long clusters, such as the marine section (light blue).

## DATA

We use synthetic cable geometries (Fig. SUPPL2) and DAS data (Figs. SUPPL3 and SUPPL4) to test the individual components of the selector, that is, spatial clustering and waveform-attribute selection. We then use seven publicly available real DAS datasets, each containing at least one recorded seismic event, to evaluate the complete ORION workflow (Figs. 2 and 3). The synthetic dataset features simple signal harmonics and full-wavefield simulations with varying levels of Gaussian noise. The real datasets span different environments: a) offshore, namely Monterey, SAFE, and CANDAS2 [32–34], b) hybrid onshore-offshore, namely Kefalonia, [35], c) urban, namely FORESEE [36], d) volcanic, namely Azuma Volcano [10], and e) geothermal areas, namely POROTOMO [37]. All datasets except Kefalonia are used to test ORION's flexibility across various geometries for a single seismic event.

The Kefalonia dataset is used for a full real-data application, with event-location accuracy serving as the evaluation metric. This dataset consists of two weeks of continuous recordings acquired from 1 to 15 August 2024 along a 15 km fiber-optic telecommunications cable connecting the islands of Kefalonia and Ithaca in Greece (Figs. 3, 4, SUPPL1a). The data recordings include waveforms from a nearby, ongoing earthquake sequence, with interevent times as short as three seconds, resulting in the detection of more than 5,700 local earthquakes using DAS data. We use a subset of 284 high-signal-to-noise-ratio (SNR) events, originally identified by [35] to build an enhanced earthquake catalog using a combination of a local network of seismometers and DAS observations. This catalog, where the integration of DAS and the local network ensures sufficient azimuthal coverage, is used as a reference to evaluate ORION's performance with DAS data within a seismic event location workflow.

Earthquake signals recorded at the Kefalonia DAS array provide an ideal test case for ORION, as the cable geometry includes both curvilinear onshore and quasi-rectilinear marine sections, with the offshore portion exhibiting significantly lower noise levels than the onshore part in the 1-30 Hz frequency band (Fig. SUPPL1b). Most earthquakes occur 10–15 km from the cable, with a few at greater distances. The azimuthal coverage is therefore limited, and events are generally more clearly recorded along the offshore segment. To efficiently locate earthquakes with spatially subsampled DAS, it is therefore essential to maximize azimuthal coverage while accounting for the low-SNR onshore channels, which are the goals of ORION.

## AUTOMATIC DAS CHANNEL SELECTION WORKFLOW

ORION is designed to spatially down-sample recordings of DAS seismic signals, thereby supporting more efficient monitoring procedures. The detected events used in this work are provided by principal investigators of the respective experiment, using different approaches, e.g., semblance-based for the Kefalonia dataset case [16].

The selection of optimal DAS channels is performed in three steps using a scoring approach: (i) subdivision of the DAS cable into subsections through automatic spatial clustering, which is based entirely on the geometry of the fiber; (ii) waveform-attribute–based selection within each spatial cluster, where attributes are converted to normalized scores; and (iii) refinement of the selected channel pool based on user inputs. In this final step, the user may adjust the weighting of data attributes by acting on score percentiles, thereby avoiding the constraint of selecting at least one channel per subsection, or, alternatively, imposing a maximum number of selected channels for each spatial cluster. While spatial clustering is performed once for a given DAS deployment, high-quality channel selection is waveform-dependent, taking into account the spatio-temporal variation in noise conditions of DAS data [38]. The following subsections describe the three steps operated by ORION.

**Spatial clustering**

DAS arrays often exhibit strong azimuthal variations over short distances. Combining channels with different azimuths can degrade signal coherence during stacking due to incompatible axial sensitivities, and signals may undergo severe degradation near abrupt changes in fiber direction. We address these issues with a modified DBSCAN algorithm [39] that enforces spatial contiguity while clustering channels by local azimuth (Figure SUPPL2). The method requires a minimum number of contiguous channels to form a cluster and excludes channels with abrupt azimuthal changes. When unclustered channels occur inside an otherwise rectilinear section, a new cluster is initiated to maintain spatial continuity. This design permits smooth azimuthal variations within a cluster, which are less problematic for stacking than sharp changes. To prevent excessively long rectilinear or gently curved sections from dominating the distribution of selected channels, we further subdivide clusters using a maximum channel count threshold. This procedure ensures balanced spatial coverage, especially for rectilinear cables or in-well deployments.

**Waveform-attributes selection**

After spatial clustering of a DAS array, we perform spatial subsampling by evaluating waveform attributes within each section for every detected event (Fig. SUPPL3). Three attributes are used for channel selection: (i) SNR, evaluated as the ratio of the mean squared amplitude of the signal window to that of the preceding noise window, expressed in decibels, (ii) local coherence, defined as the median cross-correlation coefficient between neighboring channels, and (iii) pre-event root-mean-square (RMS) trace amplitude. Together, these metrics quantify both the quality of the recorded event and the overall background noise of the channel. We keep the selector computationally efficient by choosing a few attributes, which also reduces the number of tunable parameters. At-tributes are computed over time windows defined by the event P-onset time, with onsets taken from previously-estimated picks (e.g., PhaseNet-DAS; Zhu et al., 2023). When unavailable, onsets are estimated using a conventional sliding-window SNR approach: the SNR profile is computed, smoothed, and the first prominent peak above background is identified [40]. Each attribute is normalized in the range 0-1 (with 1 representing the relative maximum SNR, MCC and minimum pre-event root-mean-square (RMS) trace amplitude) and combined into a final channel quality score. Quality scores are then used to select the best channel for each section and, for very long sections, additional channels in the subsections. Finally, to make the approach more robust against coherent noise, channels outside the 10th–90th percentile range are discarded as outliers.

**Final refinement**

After the attribute-based selection, users can further refine the pool of DAS channels by applying a lower-bound percentile threshold on the quality scores to restrict the subset to the highest-quality channels, relaxing the requirement to pick one channel per section or subsection. This refinement step may be necessary in several situations, for example, when only a few "virtual DAS stations" are desired along the DAS cable, when manual waveform inspection is planned, or when an event is clearly visible only in a subsection of the cable. The user can elect to stack channels with neighboring ones to form a "super-channel", using a default spatial window that corresponds to the gauge length; alternatively, it can also be user-defined.

**TESTS AND APPLICATIONS**

We first validate ORION using synthetic tests targeting its core components. For the spatial clustering step, we employ a complex cable geometry consisting of a rectilinear section with both abrupt and gentle bends (Fig. SUPPL2), which is used to compare the contiguity-reinforced clustering with a standard DBSCAN approach. Then, for the waveform-attribute–based channel selection and to retrieve a fixed number of traces, we use two synthetic datasets: one using simple harmonic signals and another using full-waveform seismograms [41], the latter simulating a local event recorded by a fiber cable deployed in a vertical well. Gaussian noise is added to emulate noise contamination and poorly sensitive cable sections, thereby testing ORION's capability to select high-quality DAS channels.

We apply ORION to the Kefalonia seismic sequence, targeting an event location workflow. For this goal, we use a waveform-stacking-based locator [42], adapted to handle single-component DAS data, using a local 1-D velocity model [43]. Locations using channels selected with the full ORION workflow, including stacking, are compared with those from channels uniformly sampled along the cable. To quantify location accuracy, we compute relative hypocentral distances with respect to the reference catalog of [35].

## RESULTS

### Synthetic tests

ORION identifies cable sections with coherent azimuths, similar to a standard DBSCAN. However, it additionally separates distant sections with comparable azimuths, thereby avoiding misclassification into a single cluster (Fig. SUPPL2). ORION also recovers less noise-contaminated signals across a set of 50 channels for which we have full control of the waveform attributes (Fig. SUPPL3). When applied to full-waveform data, ORION effectively avoids noisy segments and selects channels across multiple clean sections, without overrepresentation of the shallow part of the well where signals are strongest (Fig. SUPPL4).

### Real data tests

ORION identifies spatial clusters across six real-data test geometries (Fig. 2a,c,d,f), detecting abrupt (i.e., on a window length smaller than the contiguity window, which is 50 channels) azimuth variations. In portions of the cable with repetitive short-wavelength azimuth fluctuations or sharp bends (Figs. 2a-f), channels are automatically flagged as "noise" and not assigned to any spatial cluster (Figs. 2a 2b). In contrast, segments where azimuth changes occur over distances larger than the contiguity window (50 channels in these tests) are correctly grouped into single clusters (Figs. 2a, 2b). A notable case is the zig-zag configuration of the POROTOMO array (Fig. 2e), where ORION successfully resolves each branch.

With a user-defined target of 50 traces and a 90% threshold on channel quality scores, selected channels remain well distributed along the cable (Figs. 3a–d). For events recorded only weakly in certain sections (Figs. 3e, 3f), the algorithm excludes channels where events are weakly recorded, as a consequence of the percentile threshold. Finally, in very-long clusters (Figs. 3b, 3d, 3f), ORION subdivides sections into subsections to ensure sufficient spatial coverage of selected channels.

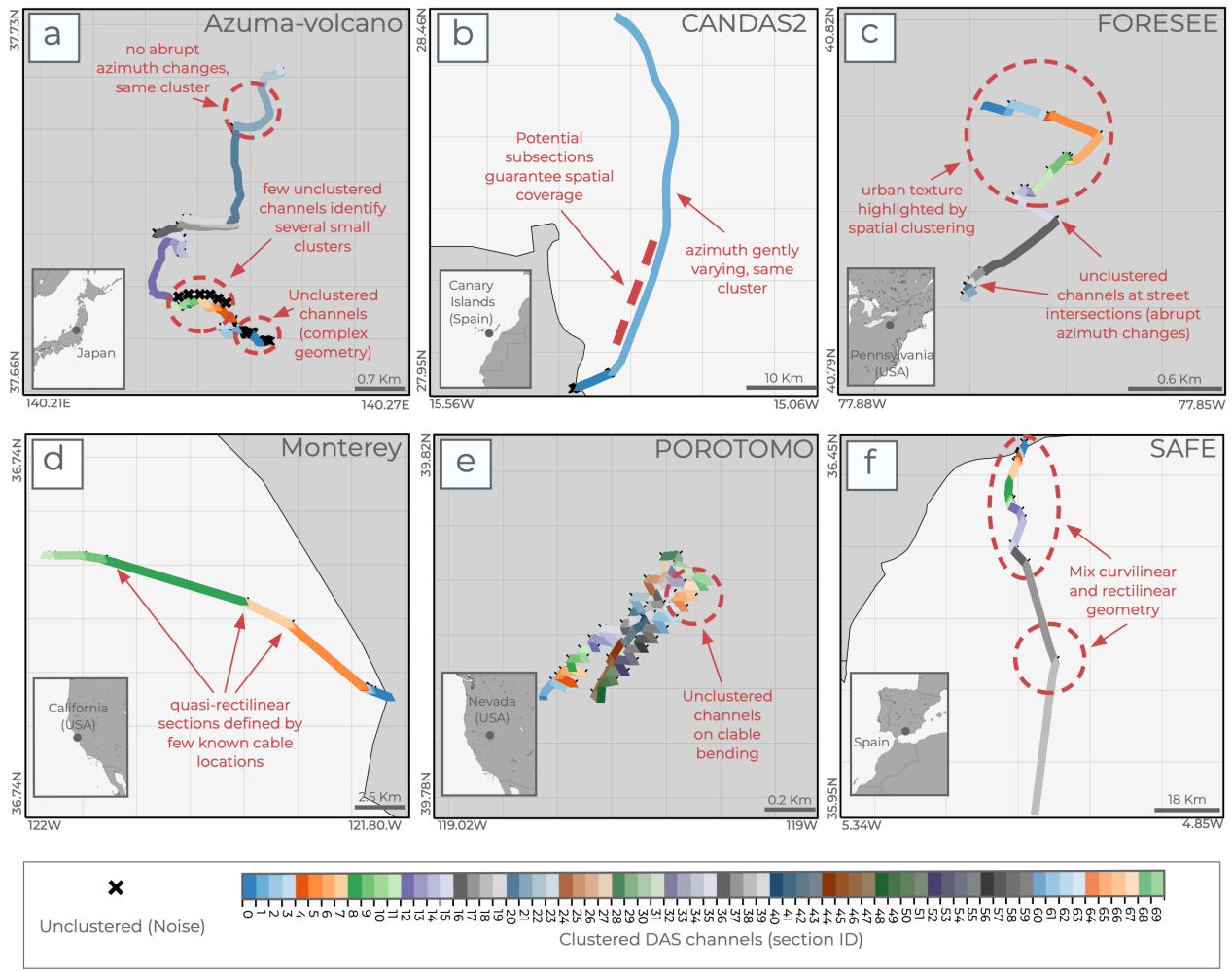

**Figure 2:** Spatial clustering with ORION tackling six different scenarios. a) Azuma-Volcano DAS deployment (Japan): The cable follows a public mountain road [10]. ORION identifies 24 clusters, mainly after the un-clustered southern section (identified as noise). The northern part, with gentler curvature, is characterized by fewer clusters. b) CANDAS-2 DAS (Spain): An extremely long-wavelength geometry results in only two clusters. c) FORESEE DAS (USA): An urban cable strictly following roads. The clustering spans all road azimuths, with unclustered channels correctly flagged at bends. d) Monterey DAS (USA): Submarine cable running below sea level. Given the weak geometrical constraints, ORION recovers clusters for the sections defined by the known location points. e) POROTOMO DAS: The zig-zag layout allows recovery of 69 clusters, with noise identified at bending points. f) SAFE DAS: Oceanic cable combining short curvilinear and long rectilinear segments, all correctly recovered by ORION.

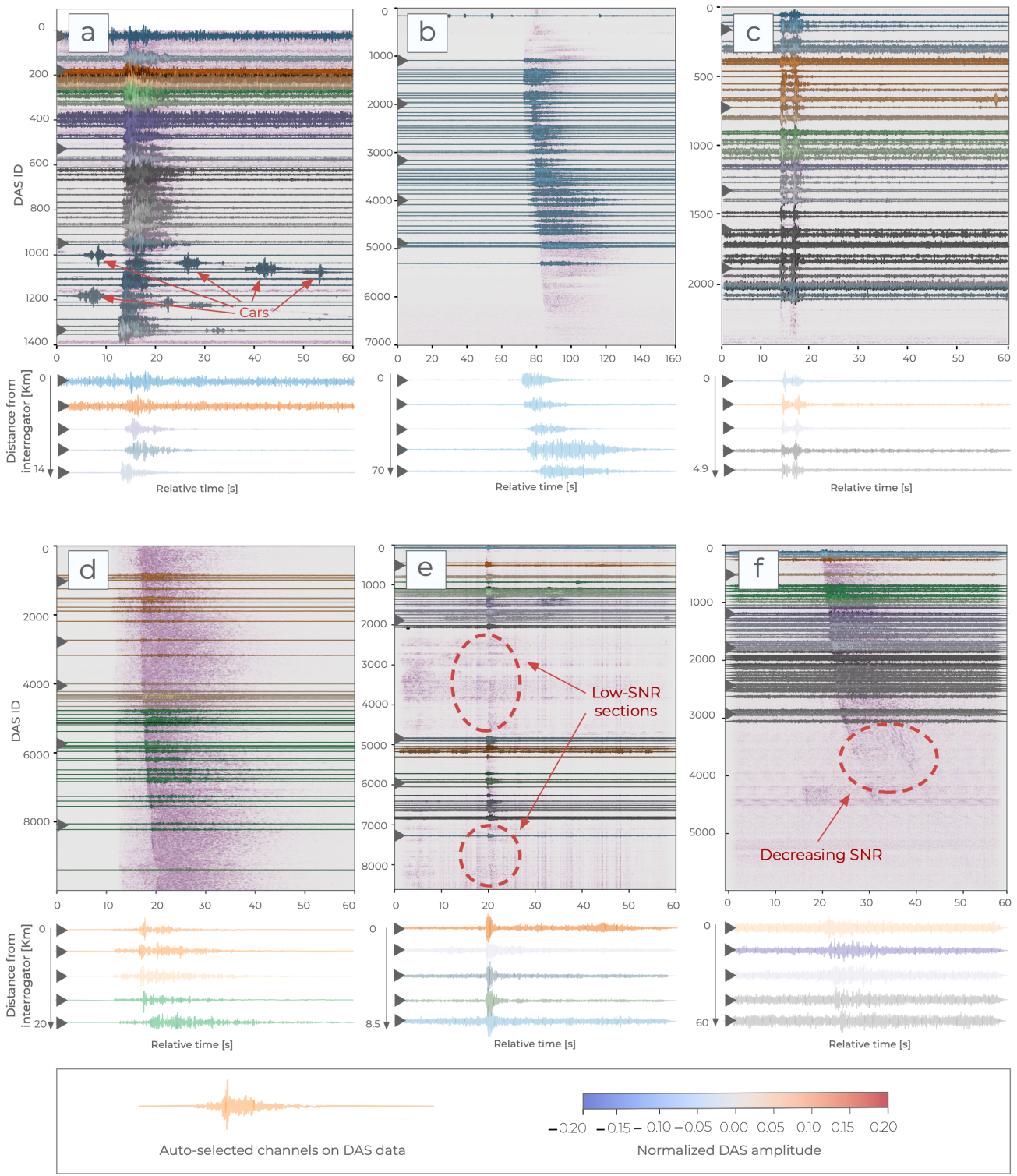

**Figure 3:** Automatic DAS channel selection using ORION. Panels refer to the datasets in Fig. 2, and colors correspond to the different spatial clusters. A 90th-percentile threshold is applied to data attribute scores; thus, not all clusters contain a selected trace (a–f). (b) A single cluster can include multiple traces if the user-defined minimum spatial coverage threshold is exceeded (subsections are identified at fixed steps). (e-f) Cable sections dominated by noise are excluded from selection due to the threshold on scores. (a-f) Additional manual selection among the automatically high-quality chosen traces is shown below the DAS data.

## Application to the Kefalonia seismic sequence

Figures 4 and 5 illustrate the application of ORION for earthquake location in the context of the 2024 seismic sequence northwest of Kefalonia island. Spatial clustering results are consistent with those from other DAS geometries: the onshore cable portions are divided into 26 segments, while the gently-varying-azimuth marine segment forms a single cluster (Figs. 4a, 4b). This marine cluster comprises roughly 2,500 channels and spans nearly 90° of azimuthal change (Fig. 4b). Nevertheless, because ORION automatically subdivides long clusters into subsections, channels with different orientations are not mixed during the subsequent stacking phase and therefore do not affect the traces used for event location.

The final set of DAS channels selected for all 284 events provides broad spatial coverage along the cable and does not overrepresent the marine segment, despite its higher data quality compared with the noisier onshore sections (Fig. 4b). Spatial coverage is further ensured by adopting a very-strict threshold on final quality scores (1st percentile), chosen to maximize azimuthal coverage for event location.

We compare event locations using ORION-selected channels with a uniform spatial selection, using the same reduced number of traces. The comparison reveals substantial differences in both signal quality and relocation performance. ORION-selected channels exhibit a higher median SNR (Fig. SUPPL5a), particularly benefiting the noisier onshore portions (Fig. SUPPL5b) and providing clearer P- and S-wave signals. When using ORION-selected channels, events cluster within the active seismicity region northwest of Kefalonia. By contrast, using uniformly selected channels, the same earthquakes tend to be mislocated near the cable (Fig. 5a). Overall, using ORION reduces the median distance to the high-resolution reference catalog from over 10 km (with uniformly spaced channels) to less than 4 km, with most mislocated events being smaller in magnitude (Fig. 5b).

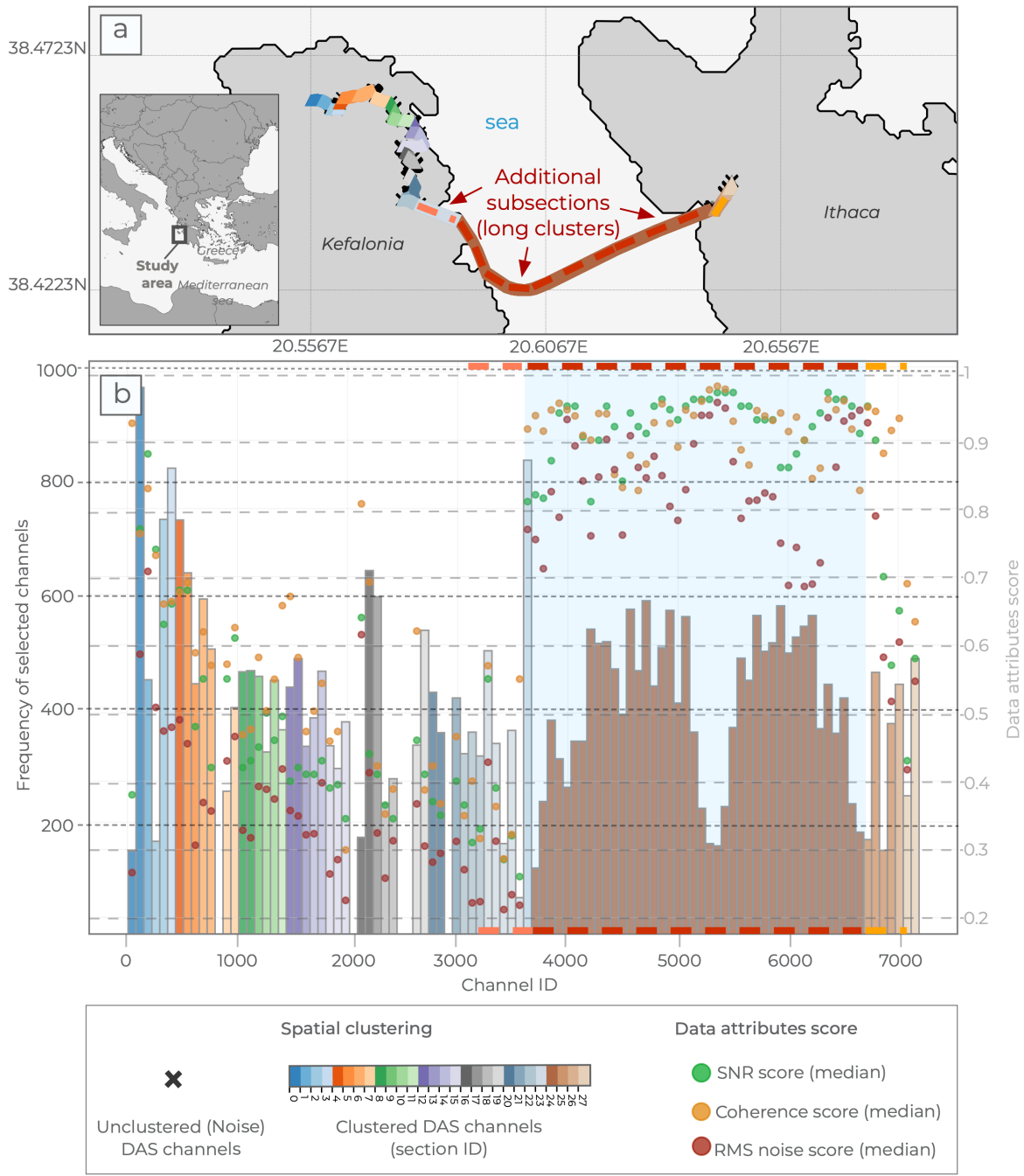

**Figure 4:** Overview of automatic DAS channel selection for 284 events from the August 2024 sequence north of Kefalonia. a) Spatial clustering by ORION identifies 27 sections, with shorter segments on land (Kefalonia and Ithaca) and the longest in the marine section. Long segments spanning multiple azimuths indicate curvature wavelengths longer than the neighbor-evaluation window. This approach accounts for successive stacking, preventing traces with sudden azimuth changes or differing sensitivities from being combined. Subsections are extracted every 50 channels to ensure full spatial coverage over those sections. b) Histogram of selected channels per cluster, showing median SNR, coherence, and pre-event noise RMS amplitudes (colored circles) used to pick optimal DAS traces. Although strong signals in the marine section (light blue) yield higher scores, ORION maintains spatial coverage, avoiding over-representation in the final selection.

**Computational costs**

ORION achieves near-real-time performance through a two-step process. First, the computationally inexpensive spatial clustering step is performed only once per DAS experiment, requiring less than one minute on an 8-CPU machine (Intel i5-11320H @ 3.20 GHz) and allowing for parameter adjustments during the experiment. The more computationally-expensive channel selection relies on three channel quality scores providing the final set of selected DAS channels within one to two minutes per event, with minimal dependence on the total channel count.

**DISCUSSION**

ORION automatically selects DAS channels in near real-time, thereby enhancing seismic monitoring analysis with fiber-optic sensors. Spatial subsampling of DAS arrays has been previously addressed either through waveform-attribute-based selection [44] or fixed-interval selection along the cable. ORION offers a complementary approach: it preserves spatial coverage using clusters of DAS channels with similar geometrical attributes, while evaluating waveform attributes to choose the optimal channels within each cluster.

Although tested on standard laptop hardware, ORION provides near real-time performance. This efficiency is possible because spatial clustering is a one-time process performed only once per cable, and the adopted waveform attributes are effective, despite being limited to SNR, local waveform coherence, and pre-event noisiness of the channel. The time required for channel selection remains negligible when compared to processing full DAS datasets, which routinely contain thousands of individual channels. Uniform selection is inherently faster because it bypasses waveform attribute evaluation; however, it fails to select the highest-quality channels, which impacts the accuracy of subsequent analysis.

ORION-selected channels produce locations that closely match reference catalog events to within 4 km, derived from both seismometers and DAS, while simultaneously reducing data volume by two orders of magnitude. The ability to define a user-specified number of traces further makes ORION suitable for hybrid seismic networks, where dense DAS arrays complement sparse conventional stations. For future permanent DAS projects, it may be useful to define a set of *super-channels*, to enable stable, long-term channel selection [45]. These super-channels could be based on the attributes explored here, but evaluated over longer periods. Super-channel identification would also help identify persistently low-quality or poorly coupled fiber sections.

## CONCLUSIONS

We present ORION, a near–real-time selector of good-quality DAS channels. ORION ensures the selection of high-quality traces while preserving sufficient spatial coverage along the cable and minimizing azimuthal gaps in subsequent seismic data analyses.

ORION is flexible across different DAS layouts, as we show its application to a range of DAS experiments without the need for extensive parameter tuning. The selection procedure runs in near real-time on a standard laptop computer, supporting expedited analysis in DAS experiments, moving beyond the typical offline procedures. Moreover, when inserted into an event location workflow, it provides more accurate hypocenter estimations compared to uniform spatial sub-sampling.

Looking forward, ORION offers potential for next-generation seismic monitoring procedures with hybrid networks, as it provides a few high-quality channels to be merged with sparse conventional networks, thereby mitigating biases that arise from differing sensor densities.

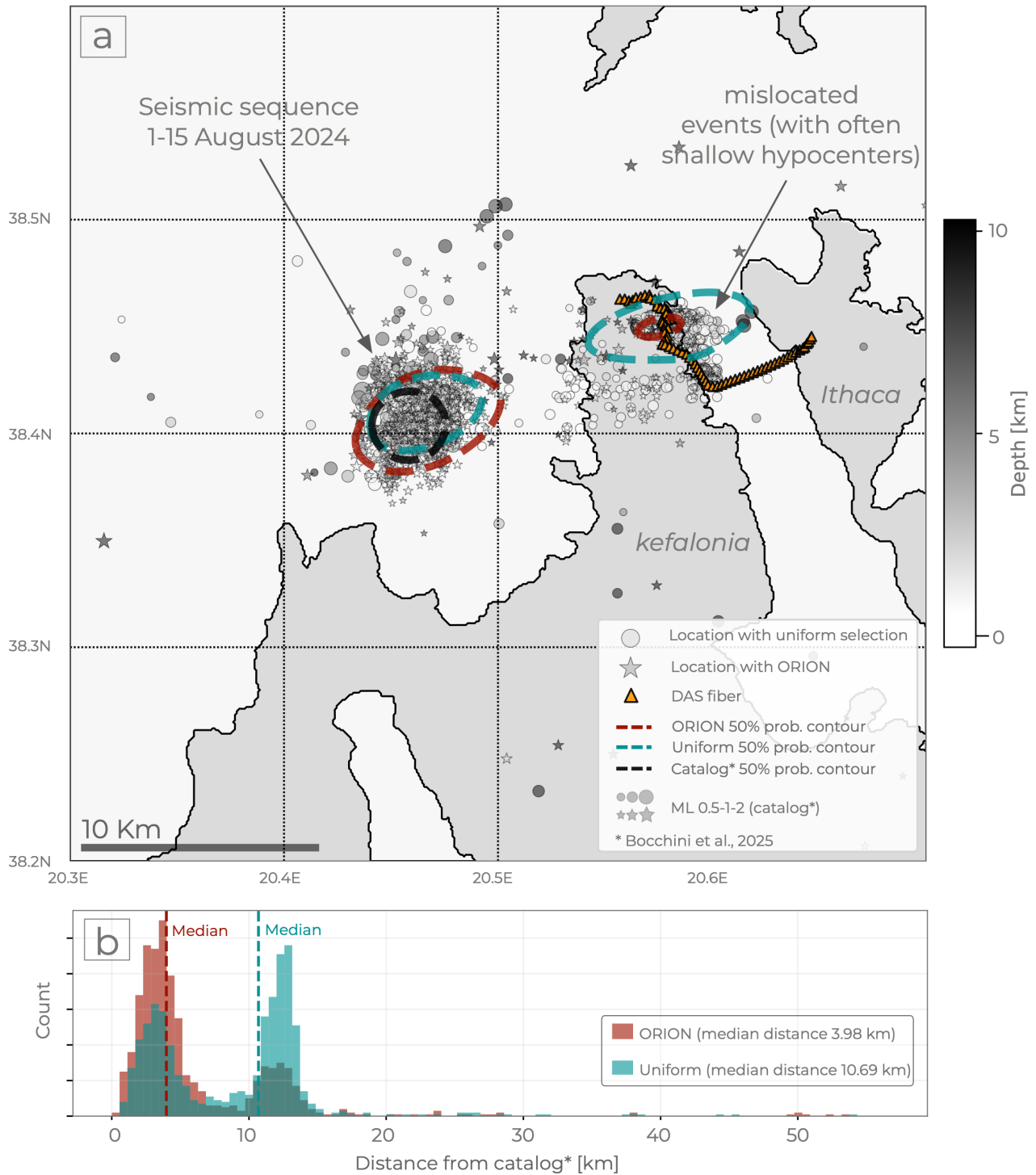

**Figure 5:** Relocation of the August 2024 Kefalonia seismic sequence using selected DAS channels and a waveform-based locator adapted for single-channel sensors [42]. Relocations use both ORION-selected and uniformly selected channels with the same number of traces (see Fig. SUPPL5). Due to the large azimuthal gap, results do not outperform the catalog obtained with both land seismometers and DAS [35], which is used as a reference. a) Event locations are concentrated in two clusters (red and blue dotted lines: 50% probability contours for ORION and uniform selection): one near the expected location north of Kefalonia and another mislocated near the cable (orange triangles). Locations with ORION-selected concentrate in the expected cluster, whereas locations with uniform selection mostly fall in the mislocated cluster. b) Histograms of distances to the reference catalog for all 284 events. Locations with ORION show a lower median distance, demonstrating improved relocation accuracy over uniform selection.

**SUBMISSION FOR PEER REVIEW**

The manuscript has been submitted for publication to Seismica.

**COMPETING INTERESTS AND AUTHOR CONTRIBUTION**

The authors declare there are no conflicts of interest for this manuscript.

Conceptualization: E.B., G.P., F.G.; Methodology: E.B., G.P., F.G.; Data curation: A.U., G.M.B., R.H.; Software: E.B., G.P.; Formal Analysis: E.B.; Visualization: E.B. ; Writing E.B.; Original draft: E.B.; Review and Editing: E.B., G.P., G.R., G.M.B., R.H., A.U., G.S., F.G.

**ACKNOWLEDGMENTS**

E.B. thanks Takeshi Nishimura, Professor at Tohoku University (Japan), for sharing the metadata of the Azuma Volcano dataset.

A.U. acknowledges support from the Severo Ochoa Centre of Excellence (grant CEX2024-001494-S) and the FERMAT project (grant PID2024-162301OB-C21) funded by MICIU/AEI/10.13039/501100011033. Canalink and Telxius facilitated fiber-optic cable access, and Aragón Photonics provided the HDAS interrogators used in the CANDAS2 and SAFE experiments, respectively.

This work has been funded by the HORIZON EU project GEOHEAT (Grant Agreement: 101147571) and GEOTWINS project (Grant Agreement: 101069750).

**DATA AVAILABILITY**

All presented DAS datasets are described in open-access papers [10, 12, 32–34, 36, 37].

The full Kefalonia DAS datasets can be found at [35].

**SOFTWARE AVAILABILITY**

The ORION Python class and data needed to reproduce the results are available in a Zenodo repository (doi:10.5281/zenodo.17464726) [46]. Additionally, future updates of the Python class will be uploaded in the following GitHub repository `https://github.com/emanuelebozzi/ORION`.

# Near real-time channel selection for Distributed Acoustic Sensing technology (Supplemental material)

**Emanuele Bozzi** ,[1] **Giulio Pascucci** ,[1] **Giacomo Rapagnani** [1] **Gian Maria Bocchini** [2] **Rebecca Harrington** [2] **Arantza Ugalde** [3] **Gilberto Saccorotti** [4] **Francesco Grigoli** [1]

[1]Department of Earth Sciences, University of Pisa, Pisa, Italy
[2]Institute of Geosciences, Ruhr University Bochum, Bochum, Germany
[3]Department of Marine Geosciences, Institut de Ciències del Mar, CSIC, 08003 Barcelona, Spain
[4]Istituto Nazionale di Geofisica e Vulcanologia, Pisa, Italy

*Corresponding author: **emanuele.bozzi@dst.unipi.it, emanuele.bozzi96@gmail.com**

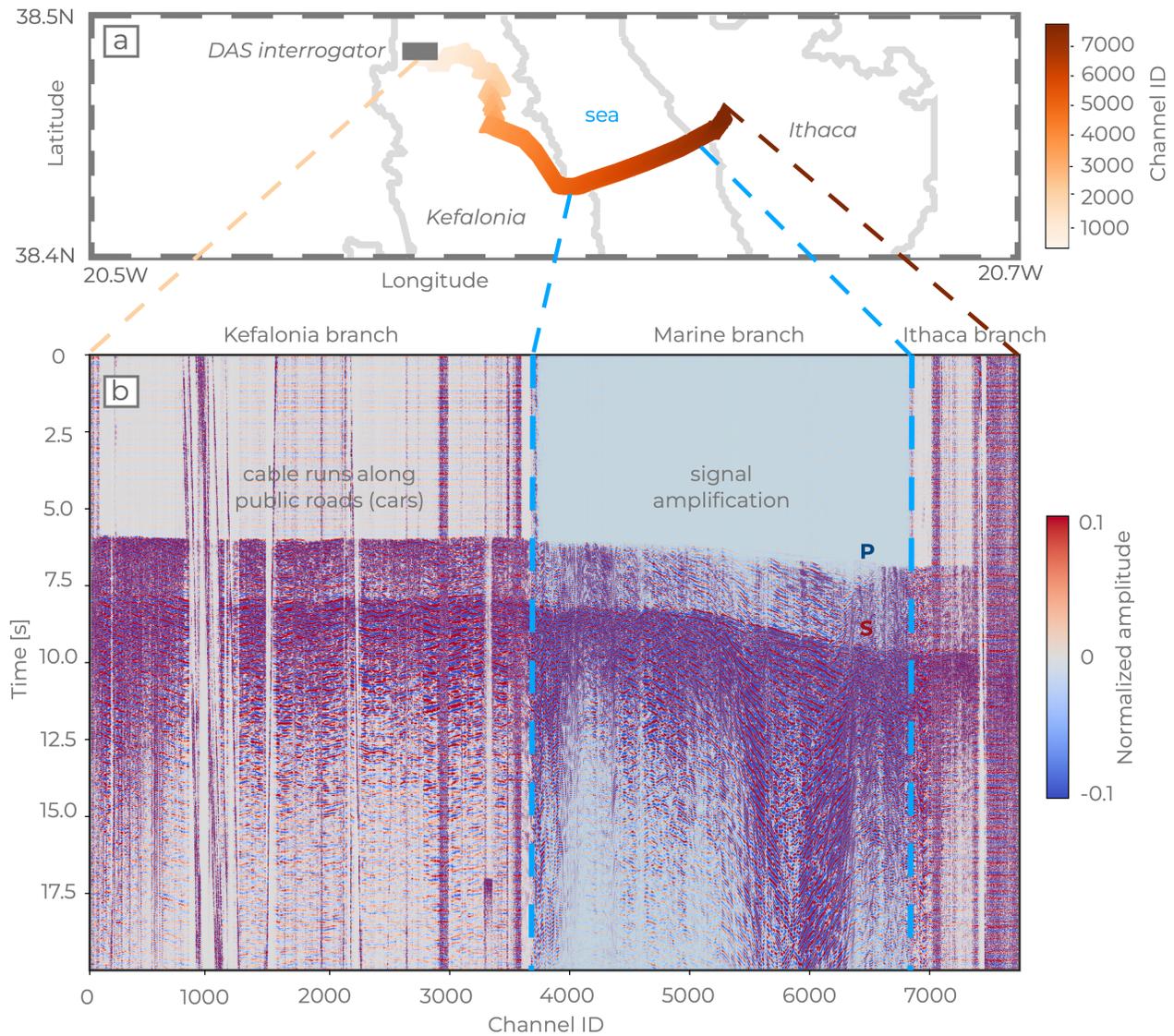

**Figure 1:** Overview on the Kefalonia DAS array used to test event relocation with selected channels from ORION. (a) Array geometry and channel layout on a simplified map of the Kefalonia-Ithaca area (see Figure 7 for a more detailed map). (b) DAS recording of an earthquake (event ID kef0367 [1]) from the target seismicity cluster north of Kefalonia Island (see Figure 7 for cluster location) recorded by the Kefalonia DAS array. Light blue dotted lines mark the marine portion of the cable, where strong signal amplification is observed. Traces are normalized individually; as a result, energetic S waves in the marine section make P waves appear less clear (see also Figure SUPPL5).

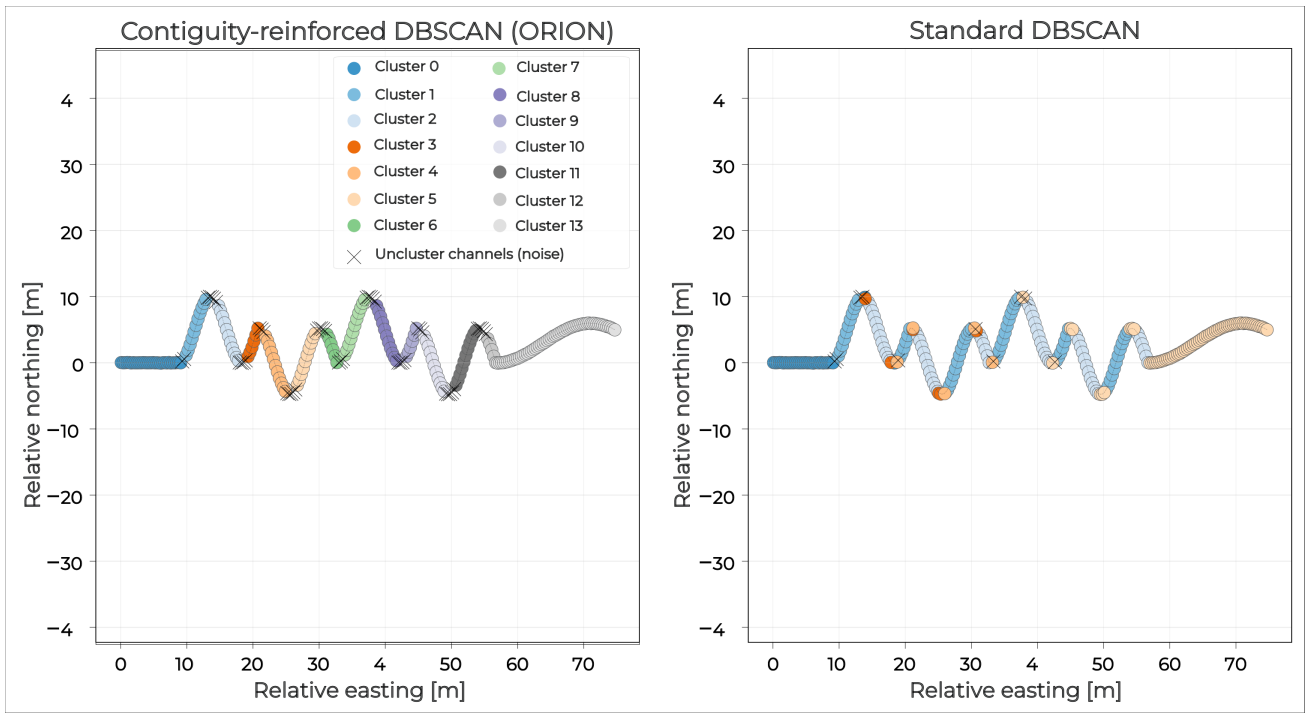

**Figure 2:** Spatial clustering of a synthetic DAS array geometry using ORION and DBSCAN [2]. Local azimuth is evaluated by the two methods, while ORION additionally evaluates spatial contiguity over a spatial window. Left) ORION identifies 13 clusters, separated by unclustered channels at each location where the cable exhibits a sudden azimuth change. Right) DBSCAN identifies 5 clusters, correctly grouping channels with similar azimuths but without distinguishing between spatially separate segments along the array.

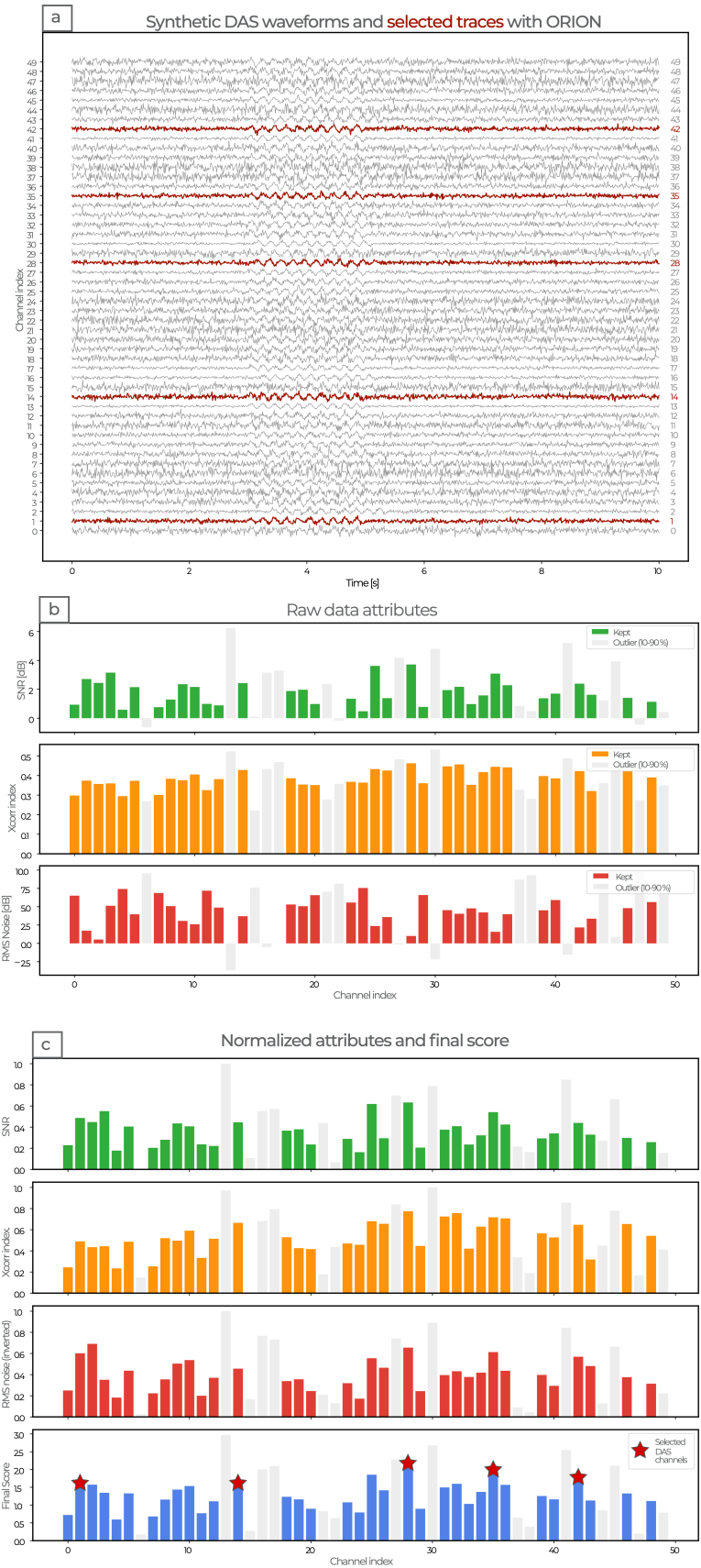

**Figure 3:** Overview of the waveform-attribute–based selection procedure in ORION. a) Synthetic waveforms simulating different levels of Gaussian noise contamination and temporal shifts of the sinusoidal signal. Red traces highlight the channels automatically selected by ORION. b) Raw waveform-attribute values for each channel, such as SNR, cross-correlation indices, and noise RMS amplitude, with outliers outside the 10th–90th percentile excluded (gray bins). c) Normalized attributes and final score values. Red stars indicate the five selected traces.

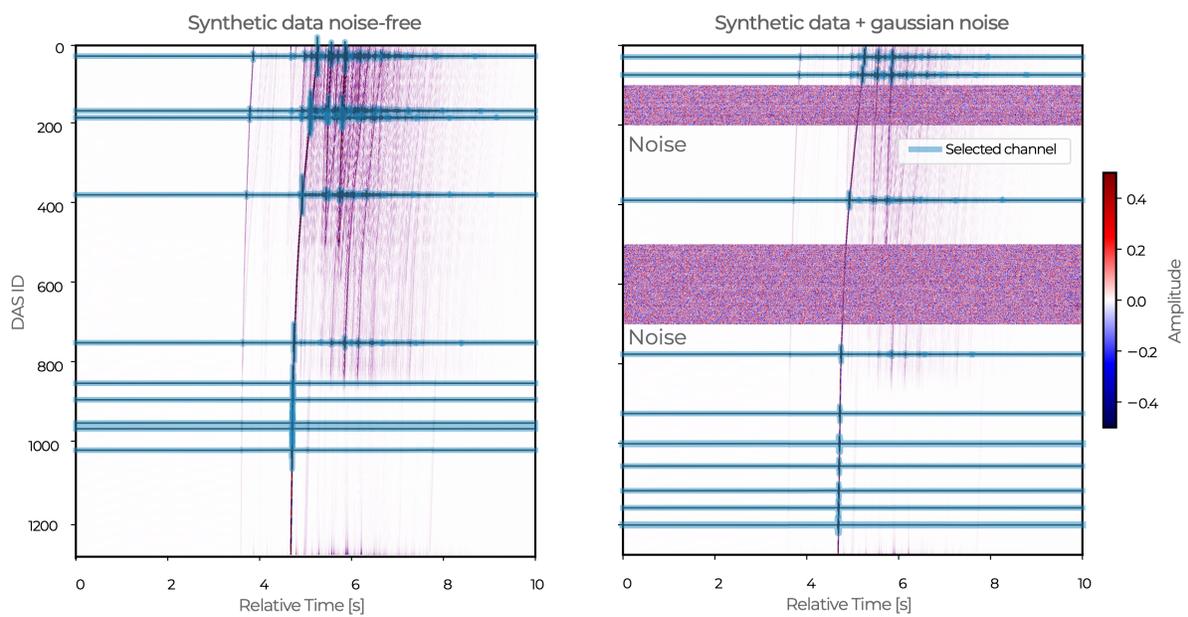

**Figure 4:** Full-waveform synthetic test of ORION. a) Noise-free synthetic DAS data generated with Salvus [3], mimicking an instrumented well and local earthquake event. The 10 traces selected by ORION are highlighted in blue. b) Noise-contaminated synthetic DAS data generated with Salvus [3], with the corresponding traces selected by ORION. Note that selected traces concentrate in noise-free areas, but are sufficiently spread along the well.

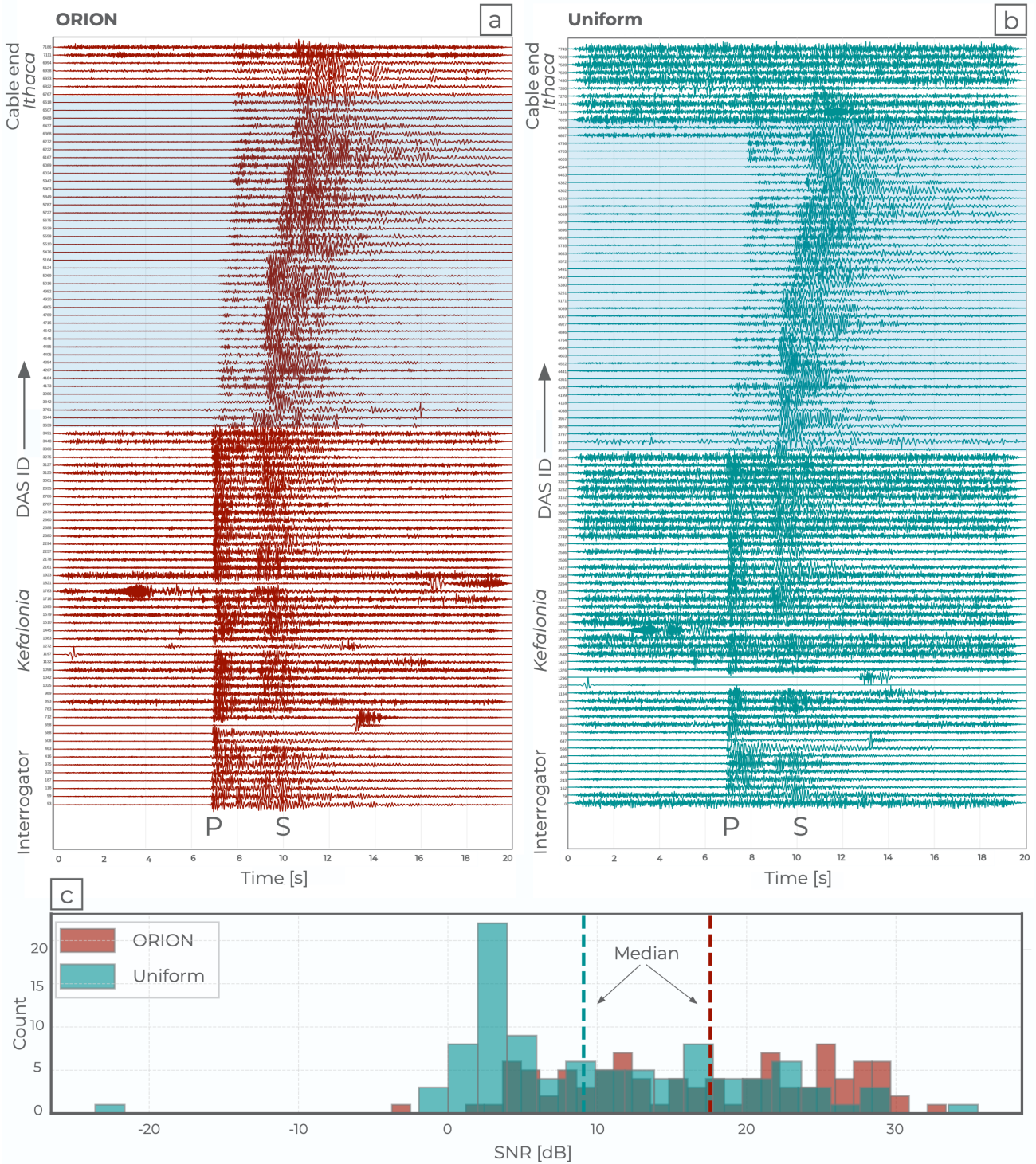

**Figure 5:** Comparison between ORION-selected DAS channels and a typical fixed-interval (uniform) selection along the cable. a) Automatic DAS channel selection for event ID kef0367 (Bocchini et al., 2025). No fixed final number of channels was imposed; instead, the minimum length of subclusters was set to 50 channels. This results in a reduction by two orders of magnitude compared with the original number of traces (over 7000). b) Selection of an identical number of traces using a fixed channel step along the cable (uniform). The selected DAS channel IDs differ between ORION and the uniform approach. c) Overall, the median SNR is higher for ORION-selected traces, with better channel quality, particularly in terms of the clarity of P- (arrival times around 6-7 s) and S-waveforms (arrival times around 10-11 s) in the land sections (Kefalonia and Ithaca).